\documentclass{article}
\usepackage{amssymb}
\usepackage{amsmath,bm}
\numberwithin{equation}{section}

\usepackage{graphicx}

\newtheorem{theorem}{Theorem}[section]

\newtheorem{remark}[theorem]{Remark}

\newcommand{\D}{\mathrm{d}}

\newcommand{\uu}{{\bf{u}}}
\newcommand{\ux}{{\bf{x}}}

\newcommand{\pp}[2]{ \frac{\partial #1}{\partial #2} }
\begin{document}
\title{ On Euler equation for incoherent fluid in curved spaces}
\author{
B. G. Konopelchenko $^{1}$  and G.Ortenzi $^{2}$ 
\footnote{Corresponding author. E-mail: giovanni.ortenzi@unito.it }\\
$^{1}$ {\footnotesize INFN, Sezione di Lecce, via Provinciale per Arnesano, 73100  Lecce, Italy } \\
 $^2$ {\footnotesize  Dipartimento di Matematica ``G. Peano'', 
Universit\`{a} di Torino, via Carlo Alberto 10, 10123, Torino, Italy}\\
$^2${\footnotesize   INFN, Sezione di Torino, via Pietro Giuria 1, 10125 Torino, Italy}
} 
\maketitle
\begin{flushright}{\footnotesize {\it To the memory of V. E. Zakharov \qquad} } \end{flushright}
\abstract{
Hodograph equations for the Euler equation in curved spaces with constant pressure are discussed. It is shown that the use of known results  
concerning geodesics and associated integrals allows to construct several types of hodograph equations. These hodograph equations
provide us with various classes of solutions of the Euler equation, including stationary solutions. Particular cases of cone and sphere in the 3-dimensional Euclidean  space are analysed in detail. Euler equation on the sphere in the 4-dimensional Euclidean space is considered too.
}

\section{Introduction}
\label{sec-intro}
Equations describing the motion of fluids and other continuous media in the curved space and space-time have attracted attention  for many years 
(see e.g. \cite{Nel80,Syn69,L-VI,GA18}).
The most simplified among them, namely, the n-dimensional  Euler equation with constant pressure, i.e. the equation
\begin{equation}
\pp{u^i}{t}+ \sum_{k=1}^n u^k \nabla_k u^i=0\, ,  \qquad i=1, \dots, n
\label{curvHEE}
\end{equation}
when $\nabla_k$ is a covariant derivative  is a substantial interest too. It describes, for example, the motion of incoherent fluid or cloud of dust in the n-dimensional curved space 
(see \cite{Syn69}).
In the present paper we study equation (\ref{curvHEE}) using the general hodograph method. The hodograph method is the classical and well-known tool to construct and study solutions of nonlinear PDEs in most cases in one dimension (see e.g. \cite{CH,L-VI,Tsa91,Whi}). Its generalization to multidimensional
case has been proposed in \cite{Zel70,Che91,Fai93} and then has been applied to construct and analyse solutions of the homogeneous Euler equation 
in dimension $n$ \cite{SZ89,FL95,Kuz03,KO22,KO-vort24,KO24}.
An extension of the hodograph method to the $n$-dimensional Euler equation with constant pressure, but with external force linear in velocity
has been discussed in the papers \cite{Che91,CC19,FKM94,KO-source24}.


Effective applicability of the general hodograph method requires the knowledge  of integrals of equations for characteristics (see e.g.  \cite{CH,Whi}). Characteristics for the
equation (\ref{curvHEE}) are the geodesics of the curved space $G$ (see e.g.  \cite{Syn69,Nel80}).
There is a number of articles devoted to the study of integrals of geodesic equation (see e.g.  \cite{BF00}).
For our purpose we need, preferably, $2n$ integrals in the $n$-dimensional space $G$ This goal can be achieved or by explicit integration of equations 
for geodesics or by the use of specific geometry of the curved space $G$.

In the present paper we study some particular cases of two- and three-dimensional spaces $G$. Surfaces of revolution in two dimensions are the best candidates since they admit two integrals in all cases.  
We analyse in detail two cases, namely, the cone and the sphere in the three-dimensional Euclidean space. We present various forms of the hodograph equations, analyse the properties of the corresponding solutions, including the conditions for blow-ups of derivatives.

The case of the sphere is rather particular, since there are three well-known integrals, namely, three components of angular momentum. This fact allows 
us to construct the particular class of the stationary solutions of equation (\ref{curvHEE}) and analyse their properties in a rather effective way.
We also study Euler equation (\ref{curvHEE}) on the 3-dimensional sphere in the 4-dimensional Euclidean space. In this case from the very beginning one
has $6$ integrals given by the components of the generators of the invariance group $SO(4)$. Using this fact one constructs the class of stationary solutions 
of the  Euler equation (\ref{curvHEE}) parametrized by two arbitrary functions of three variables.

The paper is organized as follows. General hodograph method adopted to the equation (\ref{curvHEE}) is described in Section \ref{sec-hypgeo}. Some
general formulae for the case of surfaces of revolution are given in Section \ref{sec-2rev}. Euler equation on the cone is considered in Section \ref{sec-cone}.
Hodograph equation for the Euler equation on the two-dimensional sphere are presented in section \ref{sec-S2}. Stationary solutions on the 
two-dimensional sphere are constructed and analysed in Section \ref{sec-stat-S2}. Euler equation on the 3-dimensional sphere and its stationary solutions
 is discussed in Section \ref{sec-nD}. 


\section{Integral hypersurface and geodesics}
\label{sec-hypgeo}

Here we present some well general and known facts (see e.g. \cite{CH} and \cite{KO24}) in a form adapted for our purpose.

Integral hypersurface is one of the central objects associated with the quasilinear equations in dimension $n$. It is a $(n+1)-$dimensional hypersurface in the
$(2n+1)-$dimensional space with coordinates $(t,\ux,\uu)$ defined by the system of equations
\begin{equation}
S_i(t,\ux,\uu)=0\, , \qquad i=1, \dots,n\, ,
\label{alg-sys}
\end{equation}
such that the resolution of this system with respect to $\uu$ provides us with the solution of the equation under consideration. 
Functions $S_i(t,\ux,\uu)$ obey the system of the linear equations. In the case of the Euler equation (\ref{curvHEE}) it is of the form
\begin{equation}
\pp{S_i}{t}+\sum_{k=1}^n u^k \pp{S_i}{x^k}-\sum_{m,l=1}^n \Gamma_{lm}^k u^l u^m \pp{S_i}{u^k}=0, \qquad i=1, \dots,n\, ,
\label{lin-cHEE}
\end{equation}
since the equation (\ref{curvHEE}) is equivalent to the inhomogeneous equation
\begin{equation}
\pp{u^i}{t}+\sum_{k=1}^n u^k \pp{u^i}{x^k}=-\sum_{m,l=1}^n \Gamma_{lm}^i u^l u^m, \qquad i=1, \dots,n\, ,
\label{curvHEE-expl}
\end{equation}
where $\Gamma_{lm}^i$ are Christoffel's symbols of the space $G$ with the metric $\D s^2=\sum_{m,l=1}^n g_{lm}\D x^l\D x^m$.

Any solution of the system (\ref{lin-cHEE}) provides us via (\ref{alg-sys}) with a local solution of the Euler equation (\ref{curvHEE-expl}) under the assumption
that the matrix $\partial S_i / \partial u^k$, $i,k=1, \dots,n$ is invertible. Set of $n$ arbitrary functions $\phi_i(\mathbf{S}^{(1)},\dots,\mathbf{S}^{(m)})$
of $m$ solutions $\mathbf{S}^{(i)}$ of (\ref{lin-cHEE}) is again a solution of the system (\ref{lin-cHEE}). General solution of the system (\ref{lin-cHEE})
depends on $n$ arbitrary functions of $2n$ variables. The use of such general solution in (\ref{alg-sys}) gives a general solution of equation (\ref{curvHEE}).

Method of characteristics is the standard method for construction  of solutions of the linear system (\ref{lin-cHEE}).
Characteristics for the system (\ref{lin-cHEE}) are defined by the equations
\begin{equation}
\frac{\D t}{\D \tau}=1\, , \qquad 
\frac{\D x^i}{\D \tau}=u^i\, , \qquad 
\frac{\D u^i}{\D \tau}=-\sum_{m,l=1}^n \Gamma_{lm}^iu^lu^m\, , \qquad i=1,\dots,n\, .
\label{curv-char}
\end{equation}

Hence $\tau=t$ and 
\begin{equation}
\frac{\D^2 x^i}{\D t^2}+\sum_{m,l=1}^n \Gamma_{lm}^i\frac{\D x^l}{\D t}\frac{\D x^m}{\D t}=0\, , \qquad i=1,\dots,n\, .
\label{geod-2ord}
\end{equation}
Thus, the characteristics of the system  (\ref{lin-cHEE}) are geodesics of the space $G$.
 
 Solutions $S_i$ of the system (\ref{lin-cHEE}) are constants along characteristics, i.e. 
 \begin{equation}
 \frac{\D S_i}{\D \tau}=0\, , \dots i=1,\dots,n.
 \end{equation}
So they are integrals of the dynamic system (\ref{curv-char}) or integrals for geodesics of the space $G$.

If $I_1, \dots, I_m$ are functionally independent integrals of the system (\ref{lin-cHEE}), then the functions 
\begin{equation}
S_i=\phi_i(I_1, \dots, I_m)\, , \qquad  i=1,\dots,n\, ,
\end{equation}
where  $\phi_i$ arbitrary functions, are solutions of the system (\ref{lin-cHEE}).  In this case, due to (\ref{alg-sys}), one gets solutions of the  
Euler equation (\ref{curvHEE}) depending on $n$ arbitrary functions of $m-n$ variables. One has a general solution if $m=2n$.

In virtue of equation (\ref{geod-2ord}),  the problem is reduced to the construction of integrals for the geodesic motion in the space $G$ with the
metric tensor $g_{ik}$. One of such integrals always exists for any space G: it is (see e.g. \cite{Eis26}) 
\begin{equation}
H=\sum_{i,j=1}^n g_{ij} \frac{\D x^i}{\D t}\frac{\D x^j}{\D t}=\sum_{i,j=1}^n g_{ij} u^iu^j\, .
\label{ham-gen}
\end{equation}

Construction of other integrals is a nontrivial task. It can be achieved by different methods: by defining geodesics explicitly or by use of specific geometry of the space $G$. 

Specific properties of geodesics in  curved spaces, for instance, the possibility of self-intersections, makes the property of solutions of the Euler equation
in curved space (\ref{curvHEE}) quite different from those in the flat space.

Several papers (see  \cite{CH,Syn69,Zel70, SZ89, Tsa91, Che91, CC19, Fai93, FL95, CF03, AF96, KM96, Kuz03, KM22, SZ07,  Zen11, KO22, KO24})  discuss method of characteristics, its generalizations and applications to multidimensional partial differential equations. 
In the present paper we will derive the hodograph equations for the Euler equation in curved spacetime (\ref{curvHEE}) 
and analyse some properties of its solutions.
\section{Two dimensional Euler equation on  surfaces of revolution}
\label{sec-2rev}

 In two-dimensional spaces $G$ there is a particular case for which the problem  of construction of integrals for geodesics is simplified: it is the class of surfaces of revolution. Surfaces of revolution  immersed in three-dimensional Euclidean space $\mathbb{R}^3$ with coordinates $(x,y,z)$ can be 
 parametrized using polar coordinates (see e.g. \cite{DoC})
 \begin{equation}
 x=f(\rho) \cos\phi\, , \qquad
 y=f(\rho) \sin\phi\, , \qquad
 z=h(\rho) 
 \label{para-rot}
 \end{equation}
 where $0\leq \phi <2\pi$ and $f,h$ are some functions. Metric on surface of revolution in such a parametrization is
 \begin{equation}
 \D s^2=(f'^2+h'^2)\D \rho^2 + f^2 \D\phi^2\, ,
 \end{equation}
where $f'(\rho)=\D f / \D \rho$ and  $h'(\rho)=\D h / \D \rho$\, .
 
 Christoffel symbols are 
 \begin{equation}
 \Gamma_{\rho \rho}^\rho=\frac{f' f''+h'h''}{f'^2+h'^2} \, , \qquad \Gamma_{\rho \phi}^\rho=  \Gamma_{\rho \rho}^\phi=\Gamma_{\phi \phi}^\phi= 0\, , \qquad
 \Gamma_{\phi \phi}^\rho= -\frac{f f'}{f'^2+h'^2}\, , \qquad  \Gamma_{\rho \phi}^\phi=\frac{f'}{f}\, .
 \end{equation}
Consequently, equations of geodesics are given by
\begin{equation}
\ddot{\rho}+ \frac{f' f''+h'h''}{f'^2+h'^2} \dot{\rho}^2 -\frac{f f'}{f'^2+h'^2} \dot{\phi}^2=0\, , \qquad \ddot{\phi}+ 2 \frac{f'}{f} \dot{\rho} \dot{\phi}=0\, ,
\end{equation}
where the dot indicates the time derivative.
Integral $H$ is of the form
\begin{equation}
H= (f'^2+h'^2) \dot{\rho}^2+f^2 \dot{\phi}^2\, .
\label{ene-rev}
\end{equation}
Surfaces of revolution in parametrization (\ref{para-rot}) has a particular property: they are invariant under the rotation around axes $z$. This last implies
that the component $z$ of the angular momentum 
\begin{equation}
L_z=x \dot{y}-y \dot{x}=f^2(\rho) \dot{\phi}
\label{angmomz-rev}
\end{equation}
is an integral.

So, for any surface of revolution we already have two integrals $H$ (\ref{ene-rev}) and $L_z$ (\ref{angmomz-rev}). After the standard substitution 
$\dot{\rho} \to u$, $\dot{\phi} \to v$,  one has two integrals
\begin{equation}
H= (f'^2(\rho)+h'^2(\rho)) u^2+f^2(\rho) v^2\, , \qquad L_z=f^2(\rho) v
\end{equation}
for the system (\ref{lin-cHEE}) associated with the corresponding Euler equation ($u^1=u,\, u^2=v$), i.e.
\begin{equation}
\pp{u}{t}+ u \pp{u}{\rho}+v \pp{u}{\phi} +\frac{f' f''+h'h''}{f'^2+h'^2} u^2 -\frac{f f'}{f'^2+h'^2} v^2=0\, , \qquad 
\pp{v}{t}+ u \pp{v}{\rho}+v \pp{v}{\phi}+2 \frac{f'}{f}uv=0\, .
\label{HEE-rev}
\end{equation}
Thus, it remains to find other integrals. 

Cylinder is the simplest example of surface of revolution. For cylinder
\begin{equation}
f=R=\mathit{const.}\, , \qquad  h= \rho \equiv z\, .
\end{equation}
The metric is $\D s^2=\D z^2+R^2\D\phi^2$ and all Christoffel symbols vanish. So equations of geodesics are $\ddot{z}=0$, $\ddot{\phi}=0$. 
Euler equation (\ref{HEE-rev}) becomes the homogeneous one. There are four integrals of motion $u,v$ and $\rho-ut, \phi-vt$ and 
consequently one has all the results 
already known for the 2-dimensional homogeneous Euler equation (see \cite{KO22}).
In the next two sections  we will study nontrivial particular cases.
 
\section{Euler equation on a cone.}
\label{sec-cone}
For the cone in $\mathbb{R}^3$ one has
\begin{equation}
f=\sin(\theta) r\, , \qquad h=\cos(\theta) r\,
\label{parmetcone}
\end{equation}
where $\rho=r$ and $0<\theta<\pi/2$ is a fixed angle.
One has $f'=\sin(\theta)$, $h'=\cos(\theta)$, and the metric is 
\begin{equation}
\D s^2=\D r^2+\alpha \D \phi^2 \, , \qquad \alpha \equiv \sin^2(\theta)\, .
\end{equation}
The equations of geodesics are of the form
\begin{equation}
\ddot{r}-\alpha r \dot{\phi}^2=0\, , \qquad \ddot{\phi}+\frac{2}{r}\dot{r} \dot{\phi}=0\, .
\label{geod-cone}
\end{equation}
Integrals $H$ and $L_3$ are
\begin{equation}
H=\dot{r}^2+\alpha r^2 \dot{\phi}^2\ , \qquad L_3=\alpha r^2 \dot{\phi}\, .
\label{ene-momz-cone}
\end{equation}
Note that the second equation in (\ref{geod-cone}) is equivalent to the equation $\pp{L_3}{t}=0$

So one has two independent integrals in terms of $r,u,v$
\begin{equation}
H=I_1=u^2+\alpha r^2 v^2\ , \qquad L_3=I_2=\alpha r^2 v\, .
\end{equation}
One obtains other two integrals integrating equations (\ref{geod-cone}).  First, using $\dot{\phi}=L_3/(\alpha r^2)$ and substituting it into the expansion for 
$H$, one gets
\begin{equation}
\dot{r}= \frac{\sigma}{r} \sqrt{H r^2-A}\, , 
\label{r1-cone}
\end{equation}
where $A \equiv L_3^2/\alpha>0$ and  $\sigma=\mathrm{sgn}(\dot{r}|_{t=0})$.   Integration of (\ref{r1-cone}) gives
\begin{equation}
\sqrt{H r^2-A} -\sigma Ht=\mathit{const.}\, .
\end{equation}
Fixing the constant by requiring that $r=r_0$ at $t=0$ one derives
\begin{equation}
r^2=r_0^2+Ht^2+2\sigma t \sqrt{Hr_0^2-A}\, ,
\label{solgeod-cone}
\end{equation}
or
\begin{equation}
r_0^2=r^2+Ht^2-2\sigma t \sqrt{Hr^2-A}\, .
\label{r-cone}
\end{equation}
This gives the time dependent  integral 
\begin{equation}
I_3=r^2+Ht^2-2\sigma t \sqrt{Hr^2-A}\, .
\label{I3-ene}
\end{equation}
Next, using (\ref{r-cone}) and the fact that $\dot{\phi}=L_3/(\alpha r^2)$, one gets the equation
\begin{equation}
\dot{\phi}= \frac{L_3}{\alpha} \frac{1}{r_0^2+Ht^2+2\sigma t \sqrt{Hr_0^2-A}}\, .
\end{equation}
Integrating this equation and requiring that $\phi=\phi_0$ at $t=0$, one obtains
\begin{equation}
\phi=\phi_0+\frac{L_3}{\alpha \sqrt{A}} \left( 
\mathrm{arctan} \left( \frac{Ht+\sigma \sqrt{H r_0^2-A}}{ \sqrt{A}}\right)- \mathrm{arctan} \left( \frac{\sigma \sqrt{H r_0^2-A}}{ \sqrt{A}}\right)
\right)
\end{equation}
Using (\ref{solgeod-cone}), one recover another integral
\begin{equation}
I_4= \phi_0=\phi+\frac{L_3}{\alpha \sqrt{A}} \left( 
 \mathrm{arctan} \left(\frac{\sigma  \sqrt{H r^2-A}-Ht}{ \sqrt{A}}\right)-\mathrm{arctan} \left( \frac{\sigma \sqrt{H r^2-A}}{ \sqrt{A}}\right)
\right)\, .
\label{I4-ene}
\end{equation}
Substituting the expressions for $H$ and $L_3$ given by (\ref{ene-momz-cone})  into (\ref{I3-ene}) and (\ref{I4-ene}), one gets the integrals
\begin{equation}
\begin{split}
I_3=&r^2+(u^2+\alpha r^2 v^2)t^2-2 r u t \, , \\
I_4= &\phi+\frac{1}{ \sqrt{\alpha}} \left( 
 \mathrm{arctan} \left(\frac{u r-(u^2+\alpha r^2 v)t}{  \sqrt{\alpha} r^2 v}\right)-\mathrm{arctan} \left( \frac{u}{ \sqrt{\alpha} r v}\right)
\right)\, .
\end{split}
\label{I3I4-cone}
\end{equation}
It can be  checked directly that $I_1,I_2,I_3,I_4$ are solutions of (\ref{lin-cHEE}). 

So, one has the general form of $S_i$
\begin{equation}
S_i=\Phi_i(I_1,I_2,I_3,I_4)\, , \qquad i=1,2\, ,
\end{equation}
where $\Phi_i$ are arbitrary functions.

Resolving the equations $S_i=0$ with respect to $I_3$ and $I_4$, one obtains the hodograph equations
\begin{equation}
\begin{split}
&r^2-2 r u t+(u^2+\alpha r^2 v^2)t^2=F_1(I_1,I_2) \, , \\
&\phi+\frac{1}{ \sqrt{\alpha}} \left( 
 \mathrm{arctan} \left(\frac{u r-(u^2+\alpha r^2 v)t}{  \sqrt{\alpha} r^2 v}\right)-\mathrm{arctan} \left( \frac{u}{ \sqrt{\alpha} r v}\right)
\right)=F_2(I_1,I_2) 
\end{split}
\label{hodo-cone}
\end{equation}
where $F_1$ and $F_2$ are arbitrary functions.

Differentiating equations (\ref{hodo-cone}) with respect $t$, $r$, and $\phi$, one gets the relations
\begin{equation}
M \begin{pmatrix}
\pp{u}{t} \\ \\ \pp{v}{t}
\end{pmatrix}=
 \begin{pmatrix}
A_1\\ \\ A_2
\end{pmatrix} \, , \qquad
M
 \begin{pmatrix}
\pp{u}{r} \\ \\ \pp{v}{r}
\end{pmatrix}=
 \begin{pmatrix}
B_1\\  \\B_2
\end{pmatrix} \, , \qquad
M
 \begin{pmatrix}
\pp{u}{\phi} \\ \\ \pp{v}{\phi}
\end{pmatrix}=
 \begin{pmatrix}
0\\ \\1
\end{pmatrix} \, , 
\label{der-cone}
\end{equation}
 where
 \begin{equation}
 \begin{split}
 A_1=& 2t (u^2+\alpha r^2 v^2)-2ur \, , \qquad A_2=-\frac{\alpha}{(r-ut)^2+\alpha r^2v^2t} \\
 B_1=&2r-2ut+2\alpha r v^2 t^2-2\alpha r v^2\pp{F_1}{I_1}-2\alpha r v\pp{F_1}{I_2} \, , \\
 B_2=&\frac{uvt^2}{(r-ut)^2+\alpha r^2v^2t}-2\alpha r v^2\pp{F_2}{I_1}-2\alpha r v\pp{F_2}{I_2}\, .
 \end{split}
 \end{equation}
The matrix elements  of the $2 \times 2$ matrix $M$ are
\begin{equation}
\begin{split}
M_{11}&=2u\pp{F_1}{I_1}-2ut^2+2rt\, ,\\
M_{12}&= 2 \alpha r^2 v \pp{F_1}{I_1} + \alpha r^2  \pp{F_1}{I_2}-2\alpha  r^2vt^2\, , \\
M_{21}&=2u\pp{F_2}{I_1}+ \frac{rvt^2}{(r-ut)^2+\alpha r^2v^2t}\, , \\
M_{22}&= 2\alpha r^2 v \pp{F_2}{I_1}+ \alpha r^2  \pp{F_2}{I_2} + \frac{rt(r-ut)}{(r-ut)^2+\alpha r^2v^2t} \, .
\end{split}
\end{equation}
Multiplying (\ref{der-cone}) respectively by $1$, $u$, and $v$ and using the relations
\begin{equation}
M \begin{pmatrix}
\alpha r v^2 \\  -2uv/r
\end{pmatrix}=
\begin{pmatrix}
A_1+uB_1\\ A_2+uB_2+v 
\end{pmatrix}\, ,
\end{equation}
 one gets
 \begin{equation}
M \begin{pmatrix}
\pp{u}{t}+ u \pp{u}{r} +v \pp{u}{\phi}-\alpha rv^2 \\  \\ \pp{v}{t}+ u \pp{v}{r} +v \pp{v}{\phi}+\frac{2}{r}uv
\end{pmatrix}=0 \, .
 \end{equation}
This, if $\det M \neq 0$ the variables $u$ and $v$ are indeed the solutions of the Euler equation on the cone. On the other hand the condition
$\det M = 0$ defines an hypersurface on which the derivatives blow up
\begin{equation}
\begin{split}
&\left( 2u\pp{F_1}{I_1}-2ut^2+2rt\right)\left( 2\alpha r^2 v \pp{F_2}{I_1}+ \alpha r^2  \pp{F_2}{I_2} + \frac{rt(r-ut)}{(r-ut)^2+\alpha r^2v^2t} \right)-\\
&\qquad-\left(2 \alpha r^2 v \pp{F_1}{I_1} + \alpha r^2  \pp{F_1}{I_2}-2\alpha  r^2vt^2 \right)\left( 2u\pp{F_2}{I_1}+ \frac{rvt^2}{(r-ut)^2+\alpha r^2v^2t}\right)=0\, .
\end{split}
\end{equation}

If instead of (\ref{hodo-cone}) one resolves the equations $S_1=0, \, S_2=0$ with respect to $I_1$ and $I_2$ one gets another form of hodograph equations, 
namely,
\begin{equation}
u^2+\alpha r^2v^2=\phi_1(I_3,I_4)\, , \qquad 
\alpha r^2v=\phi_2(I_3,I_4)\, , 
\end{equation}
where $\phi_1$ and $\phi_2$ are arbitrary functions. With such a choice the curve on which the derivatives blow-up is given by the equation 
\begin{equation}
\pp{\phi_1}{u}\pp{\phi_2}{v}-\pp{\phi_1}{v}\pp{\phi_2}{u}-\alpha r^2 \pp{\phi_1}{u}+2\alpha r^2 v \pp{\phi_2}{u} -2u  \pp{\phi_2}{v} +2\alpha r^2u=0\, .
\end{equation}
 In the simple case
 \begin{equation}
 \phi_1=a_1+b_1 I_3\, , \qquad  \phi_2=a_2\, , \qquad a_1,b_1,a_2 \in \mathbb{R}\, ,
 \end{equation}
one obtains
\begin{equation}
u_\pm=\frac{1}{1-b_2 t^2} \left(-b_1 r t \pm \sqrt{b_1^2r^2t^2+a_1(a-b_1t^2)-\frac{a_2^2}{\alpha r^2}(a-b_1t^2)^2} \right)\, , \qquad v=\frac{a_2}{\alpha r^2}\, .
\label{exe-cone}
\end{equation}
When $b_1>0$, the function $u_-$ blows up at $t=\pm1/\sqrt{b_1}$, while $u_+$ is regular.

At $b_1=0$ one has stationary solution of the Euler equation on the cone with dependence only on $r$, namely
\begin{equation}
u=\pm \sqrt{a_1 -\frac{a_2^2}{\alpha r^2}}\, , \qquad v=\frac{a_2}{\alpha r^2}\, .
\end{equation}

In the case $\phi_2=a_2$ and arbitrary function $\phi_1$, the function $v$ is given by (\ref{exe-cone}) while $u(t,r)$ obeys the equation
\begin{equation}
\pp{u}{t}+u \pp{u}{r}=-\frac{1}{2}\pp{}{r} \left(\frac{a_2^2}{\alpha r^2}\right)=-\pp{}{r} \left(\frac{1}{2} a_2 v\right).
\end{equation}
So, the quantity $\frac{1}{2} a_2 v$ plays the role of potential of an external force for the radial motion.

With more general choice  $\phi_1=\phi_1(I_3)$ and $\phi_2=\phi_2(I_3)$ one also gets non-stationary solutions depending only on $r$.

\section{Euler equation on a sphere $S^2$}
\label{sec-S2}

For the sphere in the 3-dimensional Euclidean space $\mathbb{R}^3$, the functions $f$ and $h$ again are given by the formulae (\ref{parmetcone}), but 
now $r=R$ is fixed and $\theta$ is a variable with range $0\leq \theta<\pi$.

The metric has the standard form 
\begin{equation}
\D s^2 = R^2 \D \theta^2 + R^2 \sin^2(\theta) \D \phi^2\, .
\end{equation}
with $0\leq \phi <2\pi$ 
and non zero Christoffel symbols are
\begin{equation}
\Gamma^\theta_{\phi \phi}=-\sin(\theta) \cos(\theta)\, , \qquad \Gamma^\phi_{\theta \phi}=	\frac{\cos(\theta)}{\sin(\theta)} \\, ,
\end{equation}

Euler equation is of the form
\begin{equation}
\begin{split}
&\pp{u}{t}+ u \pp{u}{\theta} +v \pp{u}{\phi}-\sin(\theta) \cos(\theta)v^2=0 \, ,\\
&\pp{v}{t}+ u \pp{v}{\theta} +v \pp{v}{\phi}+2\frac{\cos(\theta)}{\sin(\theta)}uv=0\, .
\end{split}
\label{HEE-sphere}
\end{equation}
 Equations of the geodesic on the sphere are given by
 \begin{equation}
 \begin{split}
 &\ddot{\theta}-\sin(\theta) \cos(\theta)\dot{\phi}^2=0\, , \\
 & \ddot{\phi}+2\frac{\cos(\theta)}{\sin(\theta)} \dot{\theta} \dot{\phi}=0\, ,
 \end{split}
\label{geod-sphere}
 \end{equation}
while integral $H$ is
\begin{equation}
H=R^2 (\dot{\theta}^2+\sin^2(\theta) \dot{\phi}^2)\, .
\label{ham-geodS2}
\end{equation}
The sphere $S^2$ is invariant under the group of rotations $SO(3)$. Consequently, all three components   $L_i=\sum_{j,k=1}^3 \epsilon_{ijk} x^j \dot{x}^k$
(where $\epsilon_{ijk}$ is the Levi-Civita antisymmetric tensor) of the angular momentum are integrals  (see e.g. \cite{BJ04}). In terms of variables 
$\theta$ and $\phi$ they are of the form
\begin{equation}
\begin{split}
L_1&= -R^2 \left(\sin(\theta) \cos(\theta) \cos(\phi) \dot{\phi}+ \sin(\phi) \dot{\theta} \right)\, ,\\
L_2&= -R^2 \left(\sin(\theta) \cos(\theta) \sin(\phi) \dot{\phi}- \cos(\phi) \dot{\theta}  \right)\, ,\\
L_3&= R^2 \sin^2(\theta)  \dot{\phi}\, ,
\end{split}
\label{mom-sphere}
\end{equation}
Note that $\mathbf{L}^2=L_1^2+L_2^2+L_3^2=R^2H$.

It is noted that components of the angular momentum evaluated on the sphere $S^2$, i.e.  the integrals (\ref{mom-sphere}) are interconnected.
Indeed, it is an easy check that for all values of $\theta$ and $\phi$  one has the relation
\begin{equation}
\cos(\phi) L_1+\sin(\phi) L_2+{\cot}(\theta) L_3=0\, .
\label{comb-mom-sphere}
\end{equation}
This relation can be obtained also  by elimination of $\dot{\theta}$ and $\dot{\phi}$ from the formulae (\ref{mom-sphere}). 

The relations  
(\ref{comb-mom-sphere}) can be viewed in another way. Indeed, let us take a point $\theta_0,\phi_0$, the  corresponding velocities 
$\dot{\theta}_0,\dot{\phi}_0$ at this point and calculate $L_1,L_2,L_3$. then, we look for the points on the sphere such that the quantities
$L_1,L_2,L_3$ have the same values as at the point $\theta_0,\phi_0$. Clearly these points should obey the relation (\ref{comb-mom-sphere})
with fixed $L_1,L_2,L_3$. 
Rewritten in the form
\begin{equation}
\cot(\theta)+a \sin(\phi+\alpha)=0
\label{geo-sphere-imp}
\end{equation}
where $a=\sqrt{L_1^2+L_2^2}/L_3$ and $\sin(\alpha)=L_1/\sqrt{L_1^2+L_2^2}$, it is the equation of great circle, i.e. geodesic in the sphere $S^2$.
We emphasize that, though the integrals $L_1,L_2,L_3$ obey the relation (\ref{comb-mom-sphere}), they remain functionally independent.

If, instead, one eliminates angles $\theta$ and $\phi$ from (\ref{mom-sphere}), one gets the constraints on $u$ and $v$, i.e.
\begin{equation}
L_1^2+L_2^2+L_3^2=R^2 \left(\dot{\theta}^2+\frac{L_3}{R^2} \dot{\phi}^2 \right)
\end{equation}
 hold on geodesics (\ref{geo-sphere-imp}).

So, in the case of sphere at the very beginning there are three functionally independent integrals. In terms of the variables $u$ and $v$ they are
\begin{equation}
\begin{split}
L_1&= -R^2 \left(\sin(\theta) \cos(\theta) \cos(\phi) v+ \sin(\phi) u \right)\, ,\\
L_2&= -R^2 \left(\sin(\theta) \cos(\theta) \sin(\phi) v- \cos(\phi) u\right)\, ,\\
L_3&= R^2 \sin^2(\theta) v\, ,
\end{split}
\end{equation}
and
\begin{equation}
H=R^2 (u^2+\sin^2(\theta) v^2)\, .
\end{equation}

In order to complete the set of integrals one needs to integrate equations of geodesics. First, using (\ref{ham-geodS2}) and the relation 
$\dot{\phi}=\frac{L_3}{R^2} \frac{1}{\sin^2(\theta)}$, one obtains the equation
\begin{equation}
\dot{\theta}=\sigma {\frac{\sqrt{H}}{R}} \frac{\sqrt{\sin^2(\theta)-k^2}}{\sin\theta}\, ,
\label{ODEene}
\end{equation}
where $\sigma=\mathrm{sgn}(u(\theta_0,\phi_0,0))$ and $k^2=\frac{L_3^2}{HR^2}<1$. Integrating (\ref{ODEene}) and requiring 
${\theta}=\theta_0$ at $t=0$, one gets
\begin{equation}
 t+ \sigma \frac{R}{\sqrt{H}} \left( \arcsin\left( \frac{\cos{\theta}}{\sqrt{1-k^2}}\right)  -\arcsin\left( \frac{\cos{\theta_0}}{\sqrt{1-k^2}}\right)   \right)=0\, .
 \label{char-theta-sphere}
\end{equation}
Hence, one has the integral
\begin{equation}
I_1= \sigma \frac{R}{\sqrt{H}} \arcsin\left( \frac{\cos{\theta_0}}{\sqrt{1-k^2}}\right)=t+ \sigma \frac{R}{\sqrt{H}} \arcsin\left( \frac{\cos{\theta}}{\sqrt{1-k^2}}\right) \, .
\end{equation}
Next, using (\ref{char-theta-sphere}) (resolved with respect to $\cos(\theta)$), one finds a solution of the equation
\begin{equation}
\dot{\phi}=\frac{L_3}{R^2}\frac{1}{\sin^2(\theta)}\, .
\end{equation}
It is given by
\begin{equation}
\begin{split}
\phi-\phi_0
=& \arctan \left( k \tan \left(  \frac{\sqrt{H}}{R}(t-I_1) \right)\right)+
\arctan \left( k \tan \left(  \frac{\sqrt{H}}{R}I_1 \right)\right) \, .
\end{split}
\end{equation}
Thus, one has the integral
\begin{equation}
I_2=\phi_0= \phi+ \arctan \left( k \tan \left( \sigma \arcsin\frac{\cos \theta}{\sqrt{1-k^2}} \right)\right)
-\arctan \left( k \tan \left(\frac{\sqrt{H}}{R}t+\sigma  \arcsin \frac{\cos \theta}{\sqrt{1-k^2}} \right)\right) \, .
\end{equation}
Since 
\begin{equation}
k=\sigma \frac{v \sin^2(\theta)}{\sqrt{u^2+\sin^2(\theta) v^2}}\, , \qquad \frac{\sqrt{H}}{R} =  \sqrt{u^2+\sin^2(\theta) v^2}\, ,
\end{equation}
one finally obtains
\begin{equation}
I_1= t+ \sigma \frac{1}{\sqrt{u^2+\sin^2(\theta) v^2}} \arcsin\left( \sqrt{\frac{u^2+\sin^2(\theta) v^2}{u^2+\sin^2(\theta)\cos^2(\theta) v^2}} \cos{\theta}\right) \, ,
\label{1-time-dep-int-S2}
\end{equation}
and
\begin{equation}
\begin{split}
I_2=& \phi+ \arctan \left( \frac{v}{u } \sin \theta \cos \theta\right)\\
&-\arctan \left( \sigma \frac{v \sin^2(\theta)}{\sqrt{u^2+\sin^2(\theta) v^2}}  \tan \left( \sqrt{u^2+\sin^2(\theta) v^2} t
+\sigma  \arcsin\left( \sqrt{\frac{u^2+\sin^2(\theta) v^2}{u^2+\sin^2(\theta)\cos^2(\theta) v^2}} \cos{\theta}\right)\right)\right)
 \, .
\end{split}
\label{2-time-dep-int-S2}
\end{equation}
 Using the properties of trigonometric functions one can rewrite $I_2$ in different equivalent forms. Thus, in the case of the sphere $S^2$ one has
 six candidates $L_1,L_2,L_3,H,I_1,I_2$ to select four independent integrals. The choice $L_3,H,I_1,I_2$ is similar to that used in the case of the cone.
 In this case one has $S_i=\Phi_i(L_3,H,I_1,I_2)$ and resolving equation (\ref{alg-sys}), one gets the hodograph equation
 \begin{equation}
 I_1=F_1(H,L_3)\, , \qquad  I_2=F_2(H,L_3)\, .
 \end{equation}
Such a choice looks also more typical for the surfaces of revolution. 

However in the case of the sphere there are other possible choices. For instance, the choice $L_1,L_2,L_3,I_1$ as four independent integrals is possible.
In such a case $S_i=\Phi_i(L_1,L_2,L_3,I_1)$  and one avoids the rather complicated integral $I_2$. Hodograph equations in this case can be chosen as

\begin{equation}
\begin{split}
 t+ \sigma \frac{1}{\sqrt{u^2+\sin^2(\theta) v^2}} \arcsin\left( \sqrt{\frac{u^2+\sin^2(\theta) v^2}{u^2+\sin^2(\theta)\cos^2(\theta) v^2}} \cos{\theta}\right)
=& F_1(L_1,L_2)\, ,\\
v \sin^2(\theta)=& F_2(L_1,L_2)\, .
\end{split}
\label{hodo-S2-simple}
\end{equation}
where $F_1$ and $F_2$ are arbitrary functions.

Taking the derivatives in $t$, $\theta$, $\phi$  of the hodograph equations in the form (\ref{hodo-S2-simple}) it is possible to obtain equations for the 
derivatives of the fields $u$ and $v$ in the form (\ref{der-cone}). The analogue of the matrix $M$ in this case is given by
\begin{equation}
M=\begin{pmatrix}
\pp{F_1}{u}-\pp{I_1}{u} &&\pp{F_1}{v}- \pp{I_1}{v} \\&&\\
 \pp{F_2}{u} &  & \pp{F_2}{v} -\sin^2(\theta)  
\end{pmatrix}
\end{equation}

The blow-up curve for the derivatives is given by the condition $ \det M= 0$, i.e.
 \begin{equation}
 \sin^2(\theta)\pp{I_1}{u}- \sin^2(\theta)\pp{F_1}{u} -\pp{I_1}{u}\pp{F_2}{v}+\pp{I_1}{v}\pp{F_2}{u}+\pp{F_1}{u}\pp{F_2}{v}-\pp{F_1}{v}\pp{F_2}{u}=0\, . 
 \end{equation}

The simplest solution of the hodograph equations (\ref{hodo-S2-simple}) corresponds to $F_1$ and $F_2$ being constants. In this case $u$ and $v$ are
independent on $\phi$  and are given by
\begin{equation}
u^2=w^2-\frac{F_2^2}{\sin^2(\theta)}\, , \qquad v=\frac{F_2}{\sin^2(\theta)}\, ,
\label{exe-spere-gen}
\end{equation}
where $w(t,\theta)$ is defined by the equation
\begin{equation}
(t-F_1)w+\sigma \arcsin\left( \sqrt{\frac{w^2}{w^2-F_2^2}} \cos\theta \right)=0\, . 
\end{equation}
The system (\ref{HEE-sphere}) is reduced to the single one-dimensional Euler equation
\begin{equation}
\pp{u}{t}+u \pp{u}{\theta}=\pp{p(\theta)}{\theta}
\label{exe-S2-1}
\end{equation}
with external force with potential
\begin{equation}
p(\theta)=-\frac{1}{2} \frac{F_2^2}{\sin^2(\theta)}=-\frac{1}{2}F_2 v\, .
\end{equation}
Solution of equation  (\ref{exe-S2-1}) are provided by the formulae (\ref{exe-spere-gen}). Various equations of type (\ref{exe-S2-1}) and their solutions
has been considered in \cite{CF03}.

 Solutions (\ref{exe-spere-gen}) blow up at the north and south poles while their derivatives blows up also on the curve given by $\pp{I_1}{u}=0$.


\section{Stationary solutions of Euler equation on sphere $\mathbf{S^2}$}
\label{sec-stat-S2}

General solutions of the Euler equation on sphere described in the previous sections correspond to the case when $S_i$, $i=1,2$ are functions of $4$ 
independent integrals. 

In Section \ref{sec-hypgeo} it was noted that in the situation when $S_i$ depend on less that $4$ integrals one gets particular subclasses of solutions. Some
of such subclasses are of interest. 

In the case of Euler equation on two-dimensional sphere one has from the very beginning three natural integrals, namely $L_1,L_2,L_3$. One gets interesting 
 solutions if $S_i$ are chosen to depend only on these three integrals, i.e.
 \begin{equation}
 S_i=\Phi_i(L_1,L_2,L_3)\, , \qquad i=1,2\, .
 \label{SstatS2}
 \end{equation}
Resolving equations $S_1=0, S_2=0$ with respect to $L_1$ and $L_2$, one obtains the following hodograph equations
\begin{equation}
\begin{split}
&\sin\theta \cos\theta \cos\phi \, v+\sin \phi \, u=F_1[\sin^2(\theta) v]\, ,\\
&\sin\theta \cos\theta \sin\phi \, v-\cos \phi \, u=F_2[\sin^2(\theta) v]\, ,
\end{split}
\label{hodo-S2-stat}
\end{equation}
where $F_1,F_2$ are arbitrary functions. So this subclass of solutions is parametrized by two arbitrary functions of single variable. Differentiating equations 
(\ref{hodo-S2-stat}) with respect to $t, \theta$, and $\phi$, one obtains
\begin{equation}
\begin{split}
&M \begin{pmatrix}
\pp{u}{t} \\ \\ \pp{v}{t}
\end{pmatrix}=
\begin{pmatrix}
0 \\ \\ 0
\end{pmatrix}\,  , \\
&M \begin{pmatrix}
\pp{u}{\theta} \\ \\ \pp{v}{\theta}
\end{pmatrix}
=\begin{pmatrix}
-v \cos\phi \cos(2 \theta) + v \sin(2 \theta) F_1'  \\ \\ -v \sin\phi \cos(2 \theta) + v \sin(2 \theta) F_2' 
\end{pmatrix}\,  , \\
&M \begin{pmatrix}
\pp{u}{\phi} \\ \\ \pp{v}{\phi}
\end{pmatrix}
=
\begin{pmatrix}
v \sin\theta \cos\theta \sin \phi -u \cos\phi\\ \\ -v \sin\theta \cos\theta \cos \phi -u \sin\phi
\end{pmatrix}\,  . 
\end{split}
\label{ders2stat}
\end{equation}
where the $2 \times 2$ matrix $M$ is
\begin{equation}
M= \begin{pmatrix}
\sin\phi& \sin\theta \cos\theta \cos\phi -\sin^2(\theta)  F_1' \\  
-\cos\phi& \sin\theta \cos\theta \sin\phi -\sin^2(\theta)  F_2'
\end{pmatrix}
\label{MstatS2}
\end{equation}
where $f'(s)=\D f(s)/\D s$.

In the case $\det M \neq 0$ the first of equations (\ref{ders2stat}) implies that $\pp{u}{t}=\pp{v}{t}=0$. So, in the case (\ref{SstatS2}) one constructs 
stationary solutions , that, in fact, is obvious from the form (\ref{hodo-S2-stat}) of the hodograph equations.

Then multiplying the second of  (\ref{ders2stat}) by $u$,  the third by $v$ summing up and using the explicit form of $M$ (\ref{MstatS2}), on gets
\begin{equation}
M
\begin{pmatrix}
u\pp{u}{\theta }+ v \pp{u}{\phi}-v^2 \sin\theta \cos\theta \\ \\
u\pp{v}{\theta }+ v \pp{v}{\phi}+2uv \frac{\cos\theta}{\sin\theta } 
\end{pmatrix}=0
\end{equation}
 So, with $\det M\neq 0$, the functions $u(\theta,\phi)$, $v(\theta,\phi)$ are indeed stationary solutions of the Euler equation on the sphere $S^2$. 
 
 This class of solutions is rather special and can be constructed explicitly. Indeed, the equations   (\ref{hodo-S2-stat}) are equivalent to the following 
\begin{equation}
\begin{split}
&\sin\theta \cos\theta  \, v-\cos\phi F_1[\sin^2(\theta) v] - \sin\phi\, F_2[\sin^2(\theta) v] =0\, ,\\
&u= \sin\phi \,F_1(\sin^2(\theta) v) - \cos \phi \,F_2[\sin^2(\theta) v] \, .
\end{split}
\label{hodo-S2-stat-red}
\end{equation}
So the problem is reduced to the resolution of a single equation given by the first equation of (\ref{hodo-S2-stat-red}). This can be done 
explicitly for a wide class of functions $F_1$ and $F_2$. 

It is noted that the solutions of the hodograph equations (\ref{hodo-S2-stat-red})  have a simple explicit dependence on the radius $R$ of the 
sphere, namely
\begin{equation}
u=\frac{1}{R^2} \tilde{u}(\theta,\phi)\, , \qquad v=\frac{1}{R^2} \tilde{v}(\theta,\phi)\, .
\end{equation}
For Euler equation (\ref{HEE-sphere}) which do not contain $R$, it is just a parameter of a certain subclass of solutions, interconnected by the scale symmetry transformation $u \to \lambda u$ and $v\to\lambda v$  ($\lambda=1/R^2$).
In order to construct stationary solutions of equation  (\ref{HEE-sphere}) independent on $R$ it is sufficient, obviously,
to take integrals $L_i/R^2$ instead of $L_i$ in formula (\ref{SstatS2}). In the rest of this section we fix $R=1$.

The simplest solution of equations (\ref{hodo-S2-stat-red}) corresponds to the linear functions $F_i$
\begin{equation}
F_1=a_1+b_1 v \sin^2(\theta)\, , \qquad  F_2=a_2+b_2 v \sin^2(\theta)\, , \qquad a_1,a_2,b_1,b_2 \in \mathbb{R}\, .
\label{fun-stat-linS2}
\end{equation}
In this case the solutions of the Euler equation are of the form
\begin{equation}
\begin{split}
u=&a_1 \sin\phi - a_2 \cos\phi + \frac{\sin\theta (a_1 \cos\phi + a_2 \sin \phi) (b_1 \sin \phi-b_2 \cos \phi)}{\cos\theta - b_1 \cos\phi \sin\theta - b_2 \sin\phi \sin\theta  } \, , \\ 
v=& \frac{a_1 \cos\phi + a_2 \sin \phi}{\sin\theta (\cos\theta - b_1 \cos\phi \sin\theta - b_2 \sin\phi \sin\theta ) }
\end{split}
\label{sol-stat-lin-S2}
\end{equation}

In the case of quadratic functions $F_1$ and $F_2$, i.e.
\begin{equation}
F_1=a_1+b_1v\sin^2(\theta)+c_1(v\sin^2(\theta))^2\, , \qquad
F_2=a_2+b_2v\sin^2(\theta)+c_2(v\sin^2(\theta))^2\, , 
\end{equation}
 one gets

\begin{equation}
{ \scriptsize
\begin{split}
u= & \sin\phi\,  (a_1+b_1v\sin^2(\theta)+c_1(v\sin^2(\theta))^2)- \cos \phi \,(a_2+b_2v\sin^2(\theta)+c_2(v\sin^2(\theta))^2) \, , \\
v=&\frac{ \cos \theta -b_1 \sin \theta\, \cos \phi\,  -b_2 \sin \theta\,    \sin \phi\,   
 \pm \sqrt{\left( \cos \theta-b_1 \sin \theta  \cos \phi\,  -b_2 \sin\theta  \sin \phi\,    \right)^2
 +4 \sin^2(\theta)\,   (a_1 \cos \phi\,  +a_2 \sin \phi\,  ) \left(c_1  \cos (\phi)+c_2\sin \phi\,  \right)} }
   {2 \sin ^3(\theta )\left(c_1 \cos \phi\,  +c_2 \sin \phi\,  \right)}
\end{split}
}
\end{equation}

Class of explicit solutions of equations (\ref{hodo-S2-stat}) is associated with the choice
\begin{equation}
F_1=a\, v \sin^2(\theta) F[v \sin^2(\theta)]\, , \qquad F_2=b\, v \sin^2(\theta) F[v \sin^2(\theta)] \, , \qquad ,a,b \in \mathbb{R}\, ,
\end{equation}
and $F$ is an arbitrary functions.
In this case the first of hodograph equations (\ref{hodo-S2-stat}) is reduced to
\begin{equation}
F[v \sin^2(\theta)]=\frac{\cot\theta}{a \cos\phi+b\sin\phi}\, .
\end{equation}
Hence, one has
\begin{equation}
\begin{split}
v=\frac{1}{\sin^2(\theta)}F^{-1}\left[\sqrt{a^2+b^2} \frac{\cot \theta}{\sin(\phi+\alpha)}  \right]\, ,\\
u=-\cot(\phi+\alpha) \cot\theta \, F^{-1}\left[\sqrt{a^2+b^2} \frac{\cot \theta}{\sin(\phi+\alpha)}  \right]\, ,
\end{split}
\end{equation}
where $a/\sqrt{a^2+b^2}=\sin\alpha$ and $F^{-1}(\xi)$ is the  function inverse to $F(\xi)$.

In particular  for
\begin{equation}
F(\xi)=d \xi^{\frac{1}{m}}\, , \qquad d,m\in \mathbb{R}\, ,
\end{equation}
one gets ($F^{-1}(\xi)=\left(\frac{\xi}{d}\right)^m$)
\begin{equation}
\begin{split}
v=&\left( \frac{a^2+b^2}{d}\right)^m \frac{1}{\sin^2(\theta)} \left( \frac{\cot\theta}{\sin(\phi+\alpha)}\right)^m\, ,\\
u=&-\left( \frac{a^2+b^2}{d}\right)^m {\cos(\phi+\alpha)} \left( \frac{\cot\theta}{\sin(\phi+\alpha)}\right)^{m+1}\, .\\
\end{split}
\label{stat-sol-S2-pol}
\end{equation}

With the choice
\begin{equation}
F(\xi)=\sqrt{\log\frac{1}{\xi}}\, ,
\end{equation}
one obtains ($F^{-1}(\xi)=e^{-\xi^2}$)\, ,
\begin{equation}
\begin{split}
v=& \frac{1}{\sin^2(\theta)} e^{-(a^2+b^2)  \frac{\cot^2\theta}{\sin^2(\phi+\alpha)}} \, ,\\
u=&  \cot(\phi+\alpha) \cot \theta \, e^{-(a^2+b^2)  \frac{\cot^2\theta}{\sin^2(\phi+\alpha)}}\, .\\
\end{split}
\label{stat-sol-S2-log}
\end{equation}
 Solutions presented above have singularities.  In particular, their behavior near poles and equator is quite simple.
 
  Indeed, near the north pole the solution (\ref{sol-stat-lin-S2}) behaves as
  \begin{equation}
  u \sim a_1 \sin \phi-a_2 \cos \phi \, , \qquad   v \sim (a_1 \cos \phi+a_2 \sin \phi) \theta^{-1}\, , \qquad \theta \to 0 \, .
  \end{equation}
Near the equator   it behaves as
\begin{equation}
\begin{split}
u&\sim a_1 \sin\phi - a_2 \cos\phi 
           + \frac{\sin\theta (a_1 \cos\phi + a_2 \sin \phi) (b_1 \sin \phi-b_2 \cos \phi)}{\cos\theta - b_1 \cos\phi \sin\theta - b_2 \sin\phi \sin\theta  }\, , \\ 
v&\sim- \frac{a_1 \cos\phi + a_2 \sin \phi}{b_1 \cos\phi - b_2 \sin\phi  }\, , \qquad \theta \to \pi/2\, .
\end{split}
\end{equation}

For the solution (\ref{stat-sol-S2-pol}) near the north pole one has
\begin{equation}
\begin{split}
u\sim &-\left( \frac{a^2+b^2}{d}\right)^m  \frac{\cos(\phi+\alpha)}{\sin^{m+1}(\phi+\alpha)}  \theta ^{-1-m}\, ,\\
v\sim &\left( \frac{a^2+b^2}{d}\right)^m   \frac{1}{\sin^m(\phi+\alpha)}  \theta ^{-2-m}\, , \qquad \theta \to 0\, .\\
\end{split}
\end{equation}
So this solution blows up at the north pole for $m>-2$ and it is finite for $m<-2$.

Near the equator  the solution behaves as
\begin{equation}
\begin{split}
u\sim &\left(- \frac{a^2+b^2}{d}\right)^m  \frac{\cos(\phi+\alpha)}{\sin^{m+1}(\phi+\alpha)}  \left(\theta-\frac{\pi}{2} \right) ^{m+1}\, ,\\
v\sim &\left( -\frac{a^2+b^2}{d}\right)^m   \frac{1}{\sin^m(\phi+\alpha)}\left(\theta-\frac{\pi}{2} \right) ^m\, , \qquad \theta \to \frac{\pi}{2}\, .\\
\end{split}
\end{equation}
So, near the equator $u$ and $v$ are finite for $m>0$ and blow up if $m<-1$.

The solution (\ref{stat-sol-S2-log}) near the north pole has the following behavior
\begin{equation}
\begin{split}
u\sim &  \cot(\phi+\alpha) \theta ^{-1}\, e^{-  \frac{(a^2+b^2)}{\theta \sin^2(\phi+\alpha)}} \to 0 \, ,\\
v\sim & \theta^{-2}e^{-\frac{(a^2+b^2)}{\theta \sin^2(\phi+\alpha)}} \to 0\, ,  \qquad \theta \to 0 \, .\\
\end{split}
\end{equation}
Near to the equator  it behaves as
\begin{equation}
\begin{split}
u\sim &  \cot(\phi+\alpha) \left(\theta-\frac{\pi}{2} \right)  \, e^{-\frac{a^2+b^2 }{\sin^2(\phi+\alpha)\left(\theta-\frac{\pi}{2} \right)^2}}\, \\
v\sim & e^{- \frac{a^2+b^2}{\sin^2(\phi+\alpha)\left(\theta-\frac{\pi}{2} \right)^2}} \, , \qquad \theta \to \frac{\pi}{2}\, .
\end{split}
\end{equation}
Solutions of the hodograph equations are regular if $\det M \neq 0$. The derivatives  blow up at the blow-up curve described by the equation
\begin{equation}
\det M = \sin\theta(\cos\theta- \sin \theta (F_1' \cos\phi+F_2' \sin\phi))=0\, .
\end{equation}
So, the derivative blows up at the north and south poles ($\theta=0,\pi/2$) for any solution and along the curve
\begin{equation}
\cos\theta- \sin \theta ( \cos\phi\,  F_1'[v(\theta,\phi) \sin^2(\theta)]+\sin\phi\, F_2'[v(\theta,\phi) \sin^2(\theta)])=0\, ,
\end{equation}
for the stationary solution $v=v(\theta,\phi)$. In the case (\ref{fun-stat-linS2}) it is the great circle 
\begin{equation}
\cot\theta -b_2 \sin\phi-b_1\cos\phi=0\, .
\label{circle-S2}
\end{equation}
Note that the solutions $u,v$ (\ref{sol-stat-lin-S2}) blow up simultaneously with their derivatives in (\ref{circle-S2}).
In the case $b_1=b_2=0$ the circle (\ref{circle-S2}) is the equator $\theta=\pi/2$.

Possible physical implications of the results presented in the section \ref{sec-S2} and \ref{sec-stat-S2} will be discussed elsewhere.

\section{Euler equation in higher dimensional spaces}
\label{sec-nD}
In dimensions $n \geq 3$ situation is more cumbersome. The cases of geodesic equations integrable in Liouville sense (see e.g. \cite{BF00}) 
are good candidates. In dimension $n$ one has in such integrable cases $n$ independent integrals from the beginning.

Other candidates are spaces of particular geometry. Indeed, similarly to the two-dimensional case, spheres in $(n+1)$-dimensional Euclidean spaces 
have very peculiar properties. 

For instance, $3$-dimensional sphere $S^3$ embedded in $4$-dimensional space is invariant under the group $SO(4)$ of rotations. Six corresponding analogs of component of angular momentum in $\mathbb{R}^3$, i.e. $x_i\dot{x}_k-\dot{x}_ix_k$, $i,k=1,2,3,4$ are integrals of geodesic flows.

The standard parametrization of the sphere in $\mathbb{R}^4$ (see e.g. \cite{Blu60}) is
\begin{equation}
\begin{split}
x^1=R \cos \phi_1\, , \quad
x^2=R \sin \phi_1 \cos \phi_2 \, , \quad
x^3=R \sin \phi_1 \sin \phi_2 \cos \phi_3\, , \quad
x^4=R \sin \phi_1 \sin \phi_2\sin \phi_3\, , 
\end{split}
\end{equation}
with $0\leq \phi_1,\phi_2 <\pi$,    $0\leq \phi_3 <2\pi$, and $R$ real positive constant. One has the metric
\begin{equation}
\D s^2=R^2 \D\phi_1^2+ R^2 \sin^2 (\phi_1) \D \phi_2^2+R^2\sin^2 (\phi_1) \sin^2 (\phi_1) \D\phi_3^2\, .
\label{metric-S3}
\end{equation}
Nonzero components of the Christoffel symbol are
\begin{equation}
\begin{split}
\Gamma^1_{22}&=-\sin\phi_1 \cos \phi_1\, , \qquad
\Gamma^1_{33}=\sin\phi_1 \cos \phi_1 \sin^2(\phi_2)\, , \\
\Gamma^2_{12}&= \cot \phi_1\, ,  \qquad
\Gamma^2_{33}= -\sin\phi_2 \cos \phi_2\, ,\\
\Gamma^3_{13}&=\cot \phi_1\, , \qquad
\Gamma^3_{23}=\cot \phi_2\, .
\end{split}
\label{Chr-S3}
\end{equation}
So the Euler equation for incoherent fluid on the sphere in $\mathbb{R}^4$ assumes the following form in the local coordinates $\phi_1,\phi_2,\phi_3$ 
\begin{equation}
\begin{split}
\pp{u^1}{t}+\sum_{k=1}^3u^k \pp{u^1}{\phi_k}=& \sin\phi_1 \cos \phi_1 (u^2)^2+ \sin\phi_1 \cos \phi_1 \sin^2(\phi_2) (u^3)^2\, , \\
\pp{u^2}{t}+\sum_{k=1}^3u^k \pp{u^2}{\phi_k}=&-2 \cot \phi_1\, u^1u^2+\sin\phi_2 \cos \phi_2 (u^3)^2 \, , \\
\pp{u^3}{t}+\sum_{k=1}^3u^k \pp{u^3}{\phi_k}=& -2 \cot \phi_1\, u^1u^2-2 \cot \phi_2\, u^2u^3\, .
\end{split}
\label{HEE-S3}
\end{equation}
Equation of geodesics are
\begin{equation}
\begin{split}
&\ddot{\phi}_1- \sin\phi_1 \cos \phi_1 (\dot{\phi}_2)^2- \sin\phi_1 \cos \phi_1 \sin^2(\phi_2) (\dot{\phi}_3)^2=0\, , \\
&\ddot{\phi}_2+2 \cot \phi_1\, \dot{\phi}_1\dot{\phi}_2-\sin\phi_2 \cos \phi_2 (\dot{\phi}_3)^2=0 \, , \\
&\ddot{\phi}_2+2 \cot \phi_1\, \dot{\phi}_1\dot{\phi}_2+2 \cot \phi_2\, \dot{\phi}_2\dot{\phi}_3=0\, .
\end{split}
\label{geod-S3}
\end{equation}
Equations (\ref{geod-S3}) have 6 integrals  given by $x^i \dot{x}^k-\dot{x}^i x^k$, $i,k=1,\dots,4$ and after the substitution $\dot{\phi}_i\to u^i$, $i=1,2,3$
these integrals are of the form
\begin{equation}
\begin{split}
L_1 =&R^2(-  \sin \phi_2 \cos \phi_3 \, u^1-\cos\phi_1\sin \phi_1  \cos \phi_2 \cos  \phi_3\, u^2
+\sin \phi_1 \cos \phi_1 \sin \phi_2  \sin \phi_3\, u^3
 ) \, ,\\
L_2 =&R^2( \cos \phi_2\,  u^1-\sin \phi_1 \cos \phi_1 \sin \phi_2  u^2) \, ,\\
L_3 =&R^2(\sin ^2(\phi_1)  \cos \phi_3\, u^2 -\sin ^2(\phi_1) \sin \phi_2 \cos \phi_2   \sin \phi_3\, u^3) \, ,\\
L_4 =&R^2(  \sin \phi_2 \sin \phi_3\, u^1 +\sin \phi_1 \cos \phi_1  \cos \phi_2 \sin \phi_3\, u^2
+\sin \phi_1 \cos \phi_1 \sin \phi_2   \cos \phi_3\, u^3) \, ,\\
L_5 =&R^2(\sin ^2(\phi_1) \sin \phi_3\, u^2+\sin ^2(\phi_1) \sin
   \phi_2 \cos \phi_2   \cos \phi_3\, u^3) \, ,\\
L_6 =&R^2(\sin ^2(\phi_1) \sin ^2(\phi_2)  u^3) \, ,
\end{split}
\label{mom-S3}
\end{equation}
Note that third geodesic equation in (\ref{geod-S3}) coincides with the condition $L_6=0$ (with $u^3=\dot{\phi}_3$). 

The integral (\ref{ham-gen}) in this case becomes 
\begin{equation}
H=\sum_{i,k=1}^3g_{ik}u^iu^k=(u^1)^2+\sin^2(\phi_1)(u^2)^2+\sin^2(\phi_1)\sin^2(\phi_2)(u^3)^2\, ,
\end{equation}
and it is related to the integrals (\ref{mom-S3}) by the relation
\begin{equation}
R^2 H=L_1^2+L_2^2+L_3^2+L_4^2+L_5^2+L_6^2\, .
\end{equation}

Similar to the two-dimensional case, six quantities $x^i \dot{x}^k-\dot{x}^i x^k$, $i,k=1,\dots,4$  evaluated on the spere $S^n$, i.e. integrals 
$L_1,\dots,L_6$ (\ref{mom-S3}) are not independent.  First let us rewrite these formulae in the following form
 \begin{equation}
 L_i=\sum_{i=1}^3 P_{ik} \dot{\phi}_k\, , \qquad 
 L_{3+i}=\sum_{i=1}^3 Q_{ik} \dot{\phi}_k\, , \qquad i=1,2,3\, ,
 \label{mom-app-S3}
 \end{equation}
where the $3\times3$ matrices $P$ and $Q$ are
\begin{equation}
P= \begin{pmatrix}
-  \sin \phi_2 \cos \phi_3  & -\cos\phi_1\sin \phi_1  \cos \phi_2 \cos  \phi_3 &
+\sin \phi_1 \cos \phi_1 \sin \phi_2  \sin \phi_3 \\
 \cos \phi_2&-\sin \phi_1 \cos \phi_1 \sin \phi_2  &0\\
0&\sin ^2(\phi_1)  \cos \phi_3 &-\sin ^2(\phi_1) \sin \phi_2 \cos \phi_2   \sin \phi_3\,
\end{pmatrix}\, ,
\label{P-mat-S3}
\end{equation}
and 
\begin{equation}
Q= \begin{pmatrix}
 \sin \phi_2 \sin \phi_3 &  +\sin \phi_1 \cos \phi_1  \cos \phi_2 \sin \phi_3 & \sin \phi_1 \cos \phi_1 \sin \phi_2   \cos \phi_3\\
0&\sin ^2(\phi_1) \sin \phi_3 & \sin ^2(\phi_1) \sin \phi_2 \cos \phi_2   \cos \phi_3\\
0&0& \sin ^2(\phi_1) \sin ^2(\phi_2)
\end{pmatrix}\, .
\label{Q-mat-S3}
\end{equation}
one observes that
\begin{equation}
\det P =0
\end{equation}
and 
\begin{equation}
\det Q =\sin^4(\phi_1)\, \sin^3(\phi_2)\, \sin^2(\phi_3)\, .
\end{equation}
Combining he relations (\ref{mom-S3}), one gets
\begin{equation}
L_i=\sum_{k=1}^3 (P Q^{-1})_{ik} L_{3+k}\, , \qquad i=1,2,3\, .
\label{comb-mom-S3}
\end{equation}
Moreover, the matrix $P$ has rank two and, consequently,
\begin{equation}
\cos\phi_2\, L_1+\sin \phi_2 \cos\phi_3\, L_2+\cot\phi_1\,L_3=0\, .
\label{comb123-mom-S3}
\end{equation}
Hence, there are two independent relations among those given bye the formula (\ref{comb-mom-S3}). So, at each point ($\phi_1,\phi_2,\phi_3$)
on the sphere $S^3$ only three integrals (\ref{mom-S3}) are linearly independent. However, all six $L_1,\dots, L_6$ are functionally independent.

As in the case of the sphere $S^2$ one can view the relations (\ref{comb-mom-S3}), (\ref{comb123-mom-S3}) in another way: consider 
(\ref{comb-mom-S3}), (\ref{comb123-mom-S3}) as the relations between the values of coordinates  $\phi_1,\phi_2,\phi_3$ for
which integrals $L_1,\dots, L_6$ have fixed constant value. There are two independent relations among those given by  (\ref{comb-mom-S3}), 
(\ref{comb123-mom-S3}) they define a curve on $S^3$.

It is noted that in the reduction to the 2-dimensional case the above formulae are reduced to those presented in the previous section. 
Indeed, under the constraint $\phi_3=0$, $u^3=0$ and the identification $\phi_1=\theta$, $\phi_2=\phi$, and $u^1=u$, $u^2=v$ the metric (\ref{metric-S3})
and the Christoffel symbols (\ref{Chr-S3}) becomes those for the 2-dimensional sphere. Equations (\ref{HEE-S3}), (\ref{geod-S3}) are reduced to
(\ref{HEE-sphere}), (\ref{geod-sphere}), integrals $L_1,L_2,L_3$ in (\ref{mom-S3})  becomes those of the formula (\ref{mom-sphere}) while integrals 
$L_4,L_5,L_6$ in (\ref{mom-S3}) vanish.
In the reduction to the sphere $S^2$ ($\phi_3=0$, $u^3=0$) the relation (\ref{comb-mom-S3}) disappear and
 the relation (\ref{comb123-mom-S3}) is reduced to (\ref{comb-mom-sphere}).

Integrals (\ref{mom-S3}) provide us with 6 functionally independent integrals required in the 3-dimensional case. 
So, it is quite natural to choose the functions $S_i$ as
\begin{equation}
S_i=\Phi_i(L_1,L_2,L_3,L_4,L_5,L_6)\, , \qquad i=1,2,3\, .
\end{equation}
 
 Resolving the equation $S_1=S_2=S_3=0$, for instance, with respect to $L_1,L_2,L_3$  one obtains the hodograph equations
 \begin{equation}
 L_{3+i}=F_i(L_1,L_2,L_3)\, , \qquad i=1,2,3\, ,
 \label{stat-hodo-S3}
 \end{equation}
where $F_i$ are arbitrary functions of $3$ variables each. Resolution of equation (\ref{stat-hodo-S3}) gives us the class of solutions $u^1,u^2,u^3$
of the Euler equation (\ref{HEE-S3}) parametrized by three arbitrary functions of two variables.

It is easy to see that these solutions are stationary similar to the two-dimensional case.

Simplest solutions corresponds to the functions $F_i$ linear in their arguments, i.e.
\begin{equation}
F_i=a_i+b_iL_1+c_iL_2+d_iL_3\, ,  \qquad a_i,b_i,c_i\in \mathbb{R}\,  \qquad i=1,2,3\, .
\end{equation}

Using the matrices $P_{ik}$ and $Q_{ik}$ defined in (\ref{P-mat-S3}) and (\ref{Q-mat-S3}) 
one presents the corresponding solution of the form 
\begin{equation}
u^k=\sum_{i=1}^3 (C^{-1})^{ki}a_i\, , \qquad k=1,2,3\, ,
\label{sol-stat-lin-S3}
\end{equation}
where $C^{-1}$ is the matrix 
\begin{equation}
C_{ik}=Q_{ik}-b_i P_{1k}-c_i Q_{2k}-d_i Q_{3k}\, , \qquad i,k=1,2,3\, .
\end{equation}
In the 2-dimensional reduction the solution (\ref{sol-stat-lin-S3}) becomes that given by the formulae (\ref{sol-stat-lin-S2}).

Other particular solutions of the hodograph equations would be of interest.

In order to find hodograph equation which will give us non-stationary solutions, one needs to find other integrals depending explicitly on time $t$, similar
to those $I_1$ and $I_2$ (\ref{1-time-dep-int-S2}) and (\ref{2-time-dep-int-S2}) in the 2-dimensional case. 

So one has to integrate the characteristic equation (\ref{geod-S3}). The use of integrals $L_i$ (\ref{mom-S3}) may simplify this task. For example, the integral
$L_6$ implies that
\begin{equation}
\dot{\phi}_3=\frac{L_6}{r^2} \frac{1}{\sin^2(\phi_1)\sin^2(\phi_2)}\, .
\end{equation}
 Using this relation, one reduces equation for geodesics (\ref{geod-S3}) to a system of equations for $\phi_1$ and $\phi_2$. The complete analysis of 
 this case will be given elsewhere.
 
 For spheres $S^n$ in the $(n+1)$-dimensional Euclidean space with $n \geq 4$, the situation is even more intriguing. Indeed, spheres $S^n$ is
 invariant under the rotation group $SO(n+1)$. The corresponding quantities $L_{ik}=x_i \dot{x}_k- \dot{x}_i x_k$, $i,k=1,\dots,n+1$ are all integrals of 
 geodesic motions. There are $\frac{n(n+1)}{2}$ of them and $\frac{n(n+1)}{2} \geq 2n$. 
 So number of functionally independent integrals exceed number of ``degrees 
 of freedom''  $2n$, which should lead to certain constraints on the geodesic motion and, consequently, on peculiar properties of solutions of the
 Euler equation.

\subsubsection*{Acknowledgments}
The authors are grateful to A. P. Veselov for useful and clarifying discussions.
This project has received funding from the European Union's Horizon 2020 research and innovation programme under the Marie Sk{\l}odowska-Curie grant no 778010 {\em IPaDEGAN}  and by the PRIN 2022TEB52W-PE1 Project ``The charm of integrability: from nonlinear waves to random matrices". We also gratefully acknowledge the auspices of the GNFM Section of INdAM, under which part of this work was carried out, and the financial support of the project MMNLP (Mathematical Methods in Non Linear Physics) of the INFN.


{}

%
%
%
%


\begin{thebibliography}{00}
\bibitem{AF96} S.I. Agafonov, E.V.Ferapontov 
{\it Systems of conservation laws in the context of the projective theory of congruences} Izv. Math. 60  1097-1122 (1996) 
\bibitem{BF00} A. V. Bolsinov, A. T. Fomenko {\it Integrable Geodesic Flows on Two-Dimensional Surfaces}, Springer US (2000)
\bibitem{BJ04} A. V. Bolsinov, B. Jovanovi\'c {\it Integrable geodesic flows on Riemannian manifolds: Construction and Obstructions} 
Contemporary Geometry And Related Topics 5-103 (2004)
\bibitem{Blu60}  L. E.  Blumenson,  {\it A Derivation of n-Dimensional Spherical Coordinates} Amer. Math. Monthly 67 (1): 63-66. (1960)
\bibitem{Che91} S. G. Chefranov {\it An exact statistical closed description of vortex turbulence and of the diffusion of an impurity in a compressible medium} Sov. Phys. Dokl. 36(4) 286-289 (1991)
\bibitem{CC19} S. G. Chefranov and A S Chefranov {\it Exact solution of the compressible Euler–Helmholtz equation and the millennium prize problem generalization}  Phys. Scr. 94 054001 (2019)
\bibitem{CH} R. Courant and D. Hilbert, {\it Methods of Mathematical Physics}, Ed. Wiley (1989)
\bibitem{CF03} T. Curtright and D. Fairlie {\it Morphing quantum mechanics and fluid dynamics} J. Phys. A: Math. Gen. 36 8885-8902 (2003)
\bibitem{DoC} M. Do Carmo, {\it Differential geometry of curves and surfaces} Prentice-Hall US (1976)
\bibitem{Eis26} L. P. Eisenhart {\it Riemannian Geometry} Princeton University Press (1997)
\bibitem{Fai93} D.B. Fairlie, {\it Equations of Hydrodynamic Type}  DTP/93/31(1993) 
\bibitem{FL95} D.B.Fairlie \& A.N.Leznov 
{\it General solutions of the Monge-Ampère equation in n-dimensional space}
J. Geom. Phys. 16(6) 385-390 (1995)
\bibitem{FKM94} G. Falkovich, E. Kuznetsov and S. Medvedev {\it Nonlinear interaction between long inertio-gravity and Rossby waves}
Nonlinear Processes in Geophysics 1 68-171 (1994)
 \bibitem{GA18} B.J. Gross, P.J. Atzberger  {\it Hydrodynamic flows on curved surfaces: Spectral numerical methods for radial manifold shapes}
J. Comp.Phys. 371 663-689 (2018)
\bibitem{KO22} B. G. Konopelchenko and G. Ortenzi {\it Homogeneous Euler equation: blow-ups, gradient catastrophes and singularity of mappings}
 J. Phys. A: Math. Theor. 55, 035203 (2022)
 \bibitem{KO-vort24} B. G. Konopelchenko and G. Ortenzi {\it On blowups of vorticity for the homogeneous Euler equation}
 SAPM  152, 5-30 (2024)
 \bibitem{KO24} B. G. Konopelchenko and G. Ortenzi  {\it On the hierarchy and fine structure of blowups and gradient catastrophes for multidimensional homogeneous Euler equation}  J. Phys. A: Math. Theor. 57  085701 (2024)
 \bibitem{KO-source24} B. G. Konopelchenko and G. Ortenzi {\it On pressureless Euler equation with external force} Physica D 469 134317 (2024)
 \bibitem{KM96}  E.A. Kuznetsov and  S.S. Minaev  {\it  Formation and propagation of cracks on the flame surface} Phys Lett A 22(1)  187- 192
 (1996)
 \bibitem{Kuz03}  E. A. Kuznetsov {\it Towards a sufficient criterion for collapse in 3D Euler equations}  Physica D 184  266-275 (2003) 
 \bibitem{KM22} E.A.Kuznetsov and  E.A. Mikhailov {\it Slipping flows and their breaking}  Ann. Phys. 447. 1690881 (2022) 
\bibitem{L-VI} L. D. Landau and E. M. Lifshitz, Fluid Mechanics, Section XV, Pergamon press (1987) 
\bibitem{Nel80} R. A. Nelson {\it Coordinate-dependent $3+1$ formulation of the general relativity equation of motion} Gen. Relat. Gravit., 12 399-404 (1980)
\bibitem{SZ07} P.M. Santini , A.I. Zenchuk, {\it The general solution of the matrix equation 
 $w_t+\sum_{k=1}^n w_{x_k}\rho^{(k)}(w)=\rho(w) + \left[ w,T \tilde{\rho}(w)\right]$} Phys  Lett A, 368  48-52  (2007)
\bibitem{SZ89}  S. F. Shandarin and Ya. B. Zel'dovich {\it The large-scale structure of the universe: Turbulence, intermittency, structures in a self-gravitating medium} Rev. Mod. Phys. 61  185-220 (1989)
\bibitem{Syn69} J.L. Synge {\it Relativity: The General Theory} Section 4, North Holland Pub. Company (1969)
\bibitem{Tsa91} S. P. Tsarev {\it The geometry of Hamiltonian systems of hydrodynamic type. The generalized hodograph method} 
Math. USSR Izv. 37  397-419 (1991)
\bibitem{Whi} G. B. Whitham {\it  Linear and Nonlinear Waves}   John Wiley \& Sons, New York, N.Y., USA (1999).
  \bibitem{Zel70}  Y. B. Zel'dovich  {\it Gravitational instability: An approximate theory for large density perturbations.}
 Astron. and Astrophys.,  5 84-89 (1970)
 \bibitem{Zen11}  A.I.Zenchuk {\it A modification of the method of characteristics:  a new class of multidimensional partially integrable 
  nonlinear systems} Phys. Lett. A, 375  2704-2713 (2011)
\end{thebibliography}
\end{document}